\documentclass{article}
\usepackage{amsmath}
\usepackage{amsfonts}

\input{tcilatex}
\begin{document}

\title{On the linear forms of the Schr\"{o}dinger equation}
\author{Y. Kasri$^{1,2}$, A. B\'{e}rard$^{2}$, Y. Grandati$^{2}$ and L.
Chetouani$^{3}$ \and $^{1}$ \textit{Laboratoire de Physique Th\'{e}orique,
Universit\'{e} Abderrahmane Mira de B\'{e}jaia,} \and \textit{Campus de
Targa Ouzemour, Dz-06000 B\'{e}jaia, Alg\'{e}rie.} \and $^{2}$\textit{\
Institut de Physique, Equipe BioPhyStat, ICPMB, IF CNRS 2843,} \and \textit{%
Universit\'{e} Paul Verlaine-Metz, 1 Bd Arago, 57078 Metz, Cedex 3, France.}
\and $^{3}$\textit{\ D\'{e}partement de Physique Th\'{e}orique, Institut de
Physique,} \and \textit{Universit\'{e} de Constantine, Route Ain El Bey,
Constantine, Algeria}.}
\date{}
\maketitle

\begin{abstract}
Generalizing the linearisation procedure used by Dirac and later by L\'{e}%
vy-Leblond, we derive the first-order non-relativistic wave equations for
particles of spin $1$ and spin $3/2$ \ starting from the Schr\"{o}dinger
equation.

By the introduction in the momentum of a correction linear in coordinates,
we establish the wave equation of the radial harmonic oscillator with
spin-orbit coupling.
\end{abstract}

\section{Introduction}

The standard procedure to derive the Dirac's free particle equation $\cite%
{Messiah}$ is based on the search of linear relation between momentum and
energy starting from Klein-Gordon equation. Making use of the Dirac idea, L%
\'{e}vy-Leblond $\cite{Levy}$ performed such linearization procedure for the
Schr\"{o}dinger equation and established a linear equation in both momentum
and energy which leads to the Pauli equation with the correct value of the
Land\'{e} $g$ factor. A generalisation of such equation was realised for
higer spin particles using the Bargmann-Wigner method in Galilean Relativity
and the Galilei-invariant wave equation $\cite{Hagen}$, $\cite{Hagen and
Hurley}$ and $\cite{Hurley}$. In $\cite{Hurley}$ Hurley has proposed, in a
rigorous and sophisticated manner, a Galilean higher spin wave equation.
Using the Galilei group and from general invariance assumptions, the author
has established a first-order wave equation with $\left( 6s+1\right) $
components for spin-s particles which admits a consistent quantum-mechanical
interpretation.

In this work, we are interested by the question of why the Dirac and L\'{e}%
vy-Leblond linearization algebraic procedure leads only to spin 1/2 wave
equation. Is it possible to generalise this method to obtain wave equations
for particles of spin greater than 1/2.

On the other hand, we focus on the non-relativistic harmonic oscillator. The
Schr\"{o}dinger's equation for this system involves a quadratic potential in
coordinates and it's seems natural in the context of first-order equations
to search a linear potential describing this system. Such problem has been
studied in relativistic quantum mechanics. The Dirac oscillator system $\cite%
{Moshinsky}$-$\cite{Kukulin}$ leads to the standard harmonic oscillator in
the non-relativistic limit, the Duffin-Kemmer-Petiau oscillator $\cite%
{Nedjadi}$-$\cite{Castro}$ is a generalisation to spin zero and one
particles. This task has been realised in the framework of the five
dimensional Galilean covariance in $\cite{de Montigny}$ and $\cite{de
Montigny 2}$.

Our purpose in this work is of double interest: the first is the
generalisation of the Dirac and L\'{e}vy-Leblond procedure to spin 1 and
spin 3/2 particles. The obtained equations are not new but coincide with
those established by Hurley. In the present paper, the derivation of the
wave equation is achieved without referring to Galilean invariance and group
theoretical techniques and it's based on the linearization concept. The
second goal is to introduce in these wave equations a linear potential in
coordinate which leads to the standard non-relativistic harmonic oscillator
plus the spin-orbit coupling for spin 1/2, 1 and 3/2 particles.

\section{Linearisation}

In the Schr\"{o}dinger equation, the presence of the energy operator $E$ is
linear while the momentum $\boldsymbol{p}$ appears quadratically in the
Hamiltonian. To eliminate this asymmetry between the time $\left( \partial
/\partial t\right) $\ and spaces variables $\left( \partial /\partial
x\right) $, we require that the wave equation must be of first order in all
derivatives. This leads to put the wave equation in the form

\begin{equation}
\left( AE+\boldsymbol{B}.\boldsymbol{p}+C\right) \psi =0  \label{1}
\end{equation}%
here $A$ , $\boldsymbol{B}\left( B_{1},B_{2},B_{3}\right) $ et $C$ is a set
of five operators with no dependence on $E$ and $\boldsymbol{p}$ and which
have to be determined. If we act on the left of $\left( \ref{1}\right) $
with the operator

\begin{equation}
A^{\prime }E-\boldsymbol{B}.\boldsymbol{p}+C^{\prime }  \label{2}
\end{equation}%
where $A^{\prime }$ and $C^{\prime }$ are a two new numerical matrices, we
find%
\begin{equation}
\left( A^{\prime }AE^{2}+\left( A^{\prime }B_{i}-B_{i}A\right) p_{i}E+\left(
A^{\prime }C+C^{\prime }A\right) E-B_{j}B_{i}p_{j}p_{i}+\left(
B_{i}C-C^{\prime }B_{i}\right) p_{i}+C^{\prime }C\right) \psi =0  \label{3}
\end{equation}%
the repeated indices are summed over $i,j=1,2,3$. If we now impose on $%
\left( \ref{3}\right) $ to be identical to Schr\"{o}dinger's equation%
\begin{equation}
\left( 2mE-\boldsymbol{p}^{2}\right) \psi =0  \label{4}
\end{equation}%
the following relations must be satisfied%
\begin{equation}
A^{\prime }A=C^{\prime }C=0,\text{ \ \ \ \ \ \ \ \ \ \ }A^{\prime
}B_{i}-B_{i}A=0,  \label{6}
\end{equation}%
\begin{equation}
A^{\prime }C+C^{\prime }A=2m,\text{ \ \ \ \ \ \ \ }C^{\prime }B_{i}-B_{i}C=0,
\label{7}
\end{equation}%
\begin{equation}
B_{j}B_{i}p_{j}p_{i}=\boldsymbol{p}^{2}.  \label{8}
\end{equation}%
Let us note that there is 7 unknown operators to be found $A$ , $B_{i}$, $C$%
, $A^{\prime }$ and $C^{\prime }$ unlike ten operators in the case of L\'{e}%
vy-Leblond procedure. We will see in the next sections that the expressions
and dimensions of these numerical matrices are related to the spin of the
particle.

\section{Particle of spin 1/2}

We start by solving the last set of algebraic equation in the spin 1/2 case
and we will reproduce the L\'{e}vy-Leblond equation by a way which could be
generalised to spin greater than one-half. The $\sigma _{i}$ Pauli matrices
obey the identity%
\begin{equation}
\sigma _{i}\sigma _{j}p_{i}p_{j}=\left( p_{i}\right) ^{2}1_{2\times 2},
\label{9}
\end{equation}%
where $1_{2\times 2}$ is a 2$\times $2 unit matrix.\ Comparing $\left( \ref%
{8}\right) $ and $\left( \ref{9}\right) $, it is straightforward to show
that $B_{j}\neq \sigma _{j}$, because otherwise the second equation in $%
\left( \ref{6}\right) $\ would lead to $A^{\prime }=A$ which contradicts the
first conditions in $\left( \ref{6}\right) $. We look for operators $B_{i}$
depending on $\sigma _{i}$ solutions of $\left( \ref{8}\right) $. Since in $%
\left( \ref{9}\right) $ there is a product of two $\sigma _{i}$ matrices
with different indices $i$ and $j$, we choose the form%
\begin{equation}
B_{i}=\left( 
\begin{array}{cc}
0 & \sigma _{i} \\ 
\sigma _{i} & 0%
\end{array}%
\right) ,  \label{14}
\end{equation}%
which leads to $\left( \ref{8}\right) $. The determination of $B_{i}$ makes
the task easier for the rest of the operators. Indeed, using $\left( \ref{6}%
\right) $ and $\left( \ref{7}\right) $, one obtains

\begin{equation}
\left( A^{\prime }+\frac{C^{\prime }}{2m}\right) B_{i}=B_{i}\left( A+\frac{C%
}{2m}\right) ,\text{ \ \ }i=1,2,3  \label{15}
\end{equation}%
\begin{equation}
\left( A^{\prime }+\frac{C^{\prime }}{2m}\right) \left( A+\frac{C}{2m}%
\right) =1,  \label{16}
\end{equation}%
here $1$\ is the unit matrix. Taking into account $\left( \ref{14}\right) $,
we see that is sufficient to put $A+\frac{C}{2m}$ equal to $1$. Combined to
the first equation in $\left( \ref{6}\right) $, this leads in one hand to $%
A^{\prime }+A=\frac{C^{\prime }}{2m}+\frac{C}{2m}=1$ and on the other hand
to $A=\frac{C^{\prime }}{2m}$ and $A^{\prime }=\frac{C}{2m}$. The following
solutions are then obtained%
\begin{equation}
A=\left( 
\begin{array}{cc}
1 & 0 \\ 
0 & 0%
\end{array}%
\right) ,C=2m\left( 
\begin{array}{cc}
0 & 0 \\ 
0 & 1%
\end{array}%
\right) .  \label{20}
\end{equation}%
The wave function $\psi =\left( 
\begin{array}{c}
\varphi \\ 
\chi%
\end{array}%
\right) $ is now a four components object where $\varphi $ and $\chi $ are
two-component spinors. Finally, the first order wave equation is written as%
\begin{equation}
\left[ \left( 
\begin{array}{cc}
1 & 0 \\ 
0 & 0%
\end{array}%
\right) E+\left( 
\begin{array}{cc}
0 & \mathbf{\sigma } \\ 
\mathbf{\sigma } & 0%
\end{array}%
\right) .\boldsymbol{p}+2m\left( 
\begin{array}{cc}
0 & 0 \\ 
0 & 1%
\end{array}%
\right) \right] \psi =0,  \label{21}
\end{equation}%
which is nothing else but the L\'{e}vy-Leblond equation reproduced here in a
different way. The Clifford algebra doesn't appear as in the original paper
of this author. However, we will see in the next sections devoted to the
spin 1 and 3/2 particles that this method can be easily generalized.

Now, we will introduce in $\left( \ref{21}\right) $ the harmonic oscillator
potential. We seek some expression that should be linear in $\boldsymbol{r}$
and leading to the three-dimensional harmonic oscillator. We choose the
substitution

\begin{equation}
\boldsymbol{p\rightarrow p}-im\omega \eta \boldsymbol{r},  \label{24.1}
\end{equation}%
where $\omega $ denote the frequency of the oscillator, $\boldsymbol{r}$ the
position vector ($\boldsymbol{r\equiv }x_{i}$) and $\eta $ is a matrix equal
to $2A^{2}-1$. The motion equation for this system becomes%
\begin{equation}
\left( AE+\boldsymbol{B}.\left( \boldsymbol{p}-im\omega \eta \boldsymbol{r}%
\right) +C\right) \psi =0,  \label{24.2}
\end{equation}%
which turns into a coupled system of equations for the components $\varphi $
and $\chi $%
\begin{eqnarray}
2m\chi &=&-\boldsymbol{\sigma }.\left( \boldsymbol{p}-im\omega \boldsymbol{r}%
\right) \varphi  \label{24.3} \\
E\varphi &=&-\boldsymbol{\sigma }.\left( \boldsymbol{p}+im\omega \boldsymbol{%
r}\right) \chi ,  \label{24.4}
\end{eqnarray}%
multiplying $\left( \ref{24.3}\right) $ by $\boldsymbol{\sigma }\left( 
\boldsymbol{p}+im\omega \boldsymbol{r}\right) $ and using $\left( \ref{24.4}%
\right) $ and the identity%
\begin{equation*}
\sigma _{i}\sigma _{j}=\delta _{ij}+i\varepsilon _{ijk}\sigma _{k}
\end{equation*}%
we obtain the wave equation for the $\varphi $ component%
\begin{equation}
E\varphi =\left[ \frac{p^{2}}{2m}+\frac{1}{2}m\omega ^{2}r^{2}-\frac{3}{2}%
\hbar \omega -\frac{2\omega }{\hbar }\left( \boldsymbol{L}.\boldsymbol{s}%
\right) \right] \varphi ,  \label{25}
\end{equation}%
where%
\begin{equation}
\boldsymbol{s}=\frac{\hbar }{2}\boldsymbol{\sigma },\text{ \ \ }\boldsymbol{%
L=r\times p}.
\end{equation}%
Equation $\left( \ref{25}\right) $ corresponds to the standard harmonic
oscillator with the addition of the spin-orbit coupling. This result is in
agreement with that obtained by the non relativistic limit of the Dirac
oscillator $\cite{Moshinsky}$.

\section{Particle of Spin 1}

The problem is to construct the five unknown operators $A$ , $B_{i}$ and $C$
obeying conditions $\left( \ref{6}\right) $ to $\left( \ref{8}\right) $\ for
a such particle. Following an approach similar to the one used above for
spin 1/2, we start by solving $\left( \ref{8}\right) $. The $B_{i}$
operators in this case, should contain the 3$\times $3 spin one matrices
which are given in the standard representation by%
\begin{equation}
\left( s_{i}\right) _{jk}=-i\hbar \varepsilon _{ijk}  \label{26}
\end{equation}%
$\varepsilon _{ijk}$ being the totally antisymmetric Levi-Cevita symbol.
Inserting the identity $-i\hbar ^{-1}\left( \boldsymbol{s}.\boldsymbol{p}%
\right) \boldsymbol{u}\mathbf{=}\boldsymbol{p}\times \boldsymbol{u}$ where $%
\boldsymbol{u=}\left( u_{1},u_{2},u_{3}\right) $ is a vector field, into the
well-known vector identity $\mathbf{\nabla }\left( \boldsymbol{\nabla }.%
\boldsymbol{u}\right) -\boldsymbol{\nabla }\times (\boldsymbol{\nabla }%
\times \boldsymbol{u}\mathbf{)=}\Delta \boldsymbol{u}$ leads to a relation
between the square of the momentum and the three components of the spin
operator, i.e.%
\begin{equation}
\hbar ^{-2}\left( N_{j}N_{i}^{T}+s_{j}s_{i}\right) p_{j}p_{i}=\boldsymbol{p}%
^{2}1_{3\times 3}  \label{28}
\end{equation}%
where $N_{i}^{T}$ denote the transposed matrix of $N_{i}$ which is a 3$%
\times $3 matrix defined by%
\begin{equation}
N_{i}=\hbar \left( 
\begin{array}{ccc}
e_{i}^{T} & 0_{3\times 1} & 0_{3\times 1}%
\end{array}%
\right)   \label{30}
\end{equation}%
with%
\begin{equation}
e_{1}=\left( 1\text{ }0\text{ }0\right) \text{, }e_{2}=\left( 0\text{ }1%
\text{ }0\right) ,e=\left( 0\text{ }0\text{ }1\right) .  \label{31}
\end{equation}%
In the same way as in the spin $1/2$ case, one may write for the $B_{i}$
matrices

\begin{equation}
B_{i}=\hbar ^{-1}\left( 
\begin{array}{ccc}
0 & N_{i} & s_{i} \\ 
N_{i}^{T} & 0 & 0 \\ 
s_{i} & 0 & 0%
\end{array}%
\right) ,  \label{32}
\end{equation}%
Thus, from $\left( \ref{32}\right) $ one obtains

\begin{equation}
B_{j}B_{i}p_{j}p_{i}=\hbar ^{-2}\left( 
\begin{array}{ccc}
N_{j}N_{i}^{T}+s_{j}s_{i} & 0 & 0 \\ 
0 & N_{j}^{T}N_{i} & N_{j}^{T}s_{i} \\ 
0 & s_{j}N_{i} & s_{j}s_{i}%
\end{array}%
\right) p_{j}p_{i}.  \label{33}
\end{equation}%
The non diagonal terms are zero%
\begin{equation}
\left( N_{j}^{T}p_{j}\right) \left( s_{i}p_{i}\right) =\left(
s_{j}p_{j}\right) \left( N_{i}p_{i}\right) =0  \label{34}
\end{equation}%
but the diagonal elements are different from zero and consequently equation $%
\left( \ref{8}\right) $ is not satisfied. This is due to the presence in $%
\left( \ref{28}\right) $ of an additional term contrary to equation $\left( %
\ref{9}\right) $ for the spin 1/2 case. Imposing in the identification
procedure that the spinor $\psi $ must be included, the condition $\left( %
\ref{8}\right) \ $becomes after acting on $\psi $ as%
\begin{equation}
B_{j}B_{i}p_{j}p_{i}\psi =p^{2}\psi .  \label{35}
\end{equation}%
In the present case, the wave function $\psi $ is a three components object
given by

\begin{equation}
\psi =\left( 
\begin{array}{c}
\boldsymbol{u} \\ 
\boldsymbol{v} \\ 
\boldsymbol{w}%
\end{array}%
\right) \text{ with }\boldsymbol{u=}\left( 
\begin{array}{c}
u_{1} \\ 
u_{2} \\ 
u_{3}%
\end{array}%
\right) ,\boldsymbol{v}=\left( 
\begin{array}{c}
v_{1} \\ 
v_{2} \\ 
v_{3}%
\end{array}%
\right) \text{ and }\boldsymbol{w=}\left( 
\begin{array}{c}
w_{1} \\ 
w_{2} \\ 
w_{3}%
\end{array}%
\right) .  \label{34.0}
\end{equation}%
It is straightforward to show that $\left( \ref{35}\right) $ is satisfied
when%
\begin{equation}
\boldsymbol{p}\mathbf{.}\boldsymbol{w}=0  \label{35.1}
\end{equation}%
\begin{equation}
\boldsymbol{v}\mathbf{\equiv }\boldsymbol{v}\left(
v_{1},v_{2}=0,v_{3}=0\right) .  \label{35.2}
\end{equation}%
This equations means that the third component of $\psi $ has a zero
divergence and that the second must be a scalar. We will see that these
conditions will be automatically fulfilled when the remaining operators $A$
and $C$ will be determined. Using $\left( \ref{32}\right) $ and in an
analogous way as in the spin one half, one can easily obtain the numerical
expressions of these matrices. Thus, the first-order free wave equation for
spin one particle takes the form%
\begin{equation}
\left[ \left( 
\begin{array}{ccc}
1 & 0 & 0 \\ 
0 & 0 & 0 \\ 
0 & 0 & 0%
\end{array}%
\right) E-i\left( 
\begin{array}{ccc}
0 & \boldsymbol{N} & \boldsymbol{s} \\ 
\boldsymbol{N}^{T} & 0 & 0 \\ 
\boldsymbol{s} & 0 & 0%
\end{array}%
\right) \boldsymbol{\nabla }+2m\left( 
\begin{array}{ccc}
0 & 0 & 0 \\ 
0 & 1 & 0 \\ 
0 & 0 & 1%
\end{array}%
\right) \right] \psi =0,  \label{39}
\end{equation}%
where $\boldsymbol{N}\mathbf{=}\left( N_{1},N_{2},N_{3}\right) $. Separating
the components, one immediately verifies the relations $\left( \ref{35.1}%
\right) $ and $\left( \ref{35.2}\right) $. Since the matrices $N_{i}^{T}$
have two rows\ composed entirely of zeros, $\left( \ref{39}\right) $ can be
reduced to an equation with $6s+1=7$ components and then coincides with the
wave equation established by Hurley $\cite{Hurley}$ using a different
approach centered on the Galilean invariance.

We now examine the harmonic oscillator problem. We introduce this potential
via the following substitution

\begin{equation}
\boldsymbol{p\rightarrow p}-im\omega \eta \boldsymbol{r,}  \label{48.0}
\end{equation}%
where the matrix $\eta $\ is given now by%
\begin{equation}
\eta =2A^{2}-1=\left( 
\begin{array}{ccc}
1 & 0 & 0 \\ 
0 & -1 & 0 \\ 
0 & 0 & -1%
\end{array}%
\right) .  \label{48.2}
\end{equation}%
Equation $\left( \ref{39}\right) $\ can be decomposed into a set of the
coupled equations%
\begin{equation}
\left\{ 
\begin{array}{c}
E\boldsymbol{u}=-\hbar ^{-1}\left[ \boldsymbol{N}\left( \boldsymbol{p}%
+im\omega \boldsymbol{r}\right) \boldsymbol{v+s}\left( \boldsymbol{p}%
+im\omega \boldsymbol{r}\right) \boldsymbol{w}\right]  \\ 
2m\boldsymbol{v}=-\hbar ^{-1}\boldsymbol{N}^{+}\left( \boldsymbol{p}%
-im\omega \boldsymbol{r}\right) \boldsymbol{u}\text{ \ \ \ \ \ \ \ \ \ \ \ \
\ \ \ \ \ \ \ \ \ \ \ } \\ 
2m\boldsymbol{w}=-\hbar ^{-1}\boldsymbol{s}\left( \boldsymbol{p}-im\omega 
\boldsymbol{r}\right) \boldsymbol{u,}\text{\ \ \ \ \ \ \ \ \ \ \ \ \ \ \ \ \
\ \ \ \ \ \ \ \ \ }%
\end{array}%
\right. 
\end{equation}%
thus the wave equation for the field $\boldsymbol{u}$ is given by%
\begin{equation}
2mE\boldsymbol{u}=\hbar ^{-2}\left[ \left( \boldsymbol{N}.\boldsymbol{p}%
_{+}\right) \left( \boldsymbol{N}^{+}.\boldsymbol{p}_{-}\right) +\left( 
\boldsymbol{s}.\boldsymbol{p}_{+}\right) \left( \boldsymbol{s}.\boldsymbol{p}%
_{-}\right) \right] \boldsymbol{u}  \label{48.3}
\end{equation}%
where $\boldsymbol{p}_{\pm }=\boldsymbol{p}\pm im\omega \boldsymbol{r}$.
From the definitions $\left( \ref{26}\right) $ and $\left( \ref{30}\right) $%
, one readily shows that $N_{i}$ and $s_{i}$ verify%
\begin{equation}
N_{i}N_{j}^{+}+s_{i}s_{j}=i\hbar \varepsilon _{ijk}s_{k}+\hbar ^{2}\delta
_{ij}.  \label{49}
\end{equation}%
With the help of the above relation, the evaluation of $\left( \ref{48.3}%
\right) $ yields

\begin{equation}
E\boldsymbol{u}=\left[ \frac{p^{2}}{2m}+\frac{1}{2}m\omega ^{2}r^{2}-\frac{3%
}{2}\hbar \omega -\frac{\omega }{\hbar }\left( \boldsymbol{L}.\boldsymbol{s}%
\right) \right] \boldsymbol{u}  \label{48.10}
\end{equation}%
which describes the usual isotropic harmonic oscillator plus the spin-orbit
interaction with the strength $\left( -\omega /\hbar \right) $. Let us note
that this strength is one half that obtained for spin 1/2 case and it's
coincides with the non relativistic limit of the DKP oscillator $\cite%
{Nedjadi}$ and that obtained in the framework of the five dimensional
Galilean covarience $\cite{de Montigny 2}$.

\section{Spin 3/2 particle}

For such particle, we will adopt the following matrix representation for the
components of spin operator $\boldsymbol{s}$

\begin{eqnarray}
s_{1} &=&\hbar \left( 
\begin{array}{cccc}
0 & \frac{1}{2}\sqrt{3} & 0 & 0 \\ 
\frac{1}{2}\sqrt{3} & 0 & 1 & 0 \\ 
0 & 1 & 0 & \frac{1}{2}\sqrt{3} \\ 
0 & 0 & \frac{1}{2}\sqrt{3} & 0%
\end{array}%
\right) ,s_{2}=\hbar \left( 
\begin{array}{cccc}
0 & -i\frac{1}{2}\sqrt{3} & 0 & 0 \\ 
i\frac{1}{2}\sqrt{3} & 0 & -i & 0 \\ 
0 & i & 0 & -i\frac{1}{2}\sqrt{3} \\ 
0 & 0 & i\frac{1}{2}\sqrt{3} & 0%
\end{array}%
\right) ,  \label{50} \\
s_{3} &=&\hbar \left( 
\begin{array}{cccc}
\frac{3}{2} & 0 & 0 & 0 \\ 
0 & \frac{1}{2} & 0 & 0 \\ 
0 & 0 & -\frac{1}{2} & 0 \\ 
0 & 0 & 0 & -\frac{3}{2}%
\end{array}%
\right) .  \notag
\end{eqnarray}

We seek the $B_{i}$ solutions of $\left( \ref{8}\right) $\ containing the
hermitian matrices $s_{i}$. We introduce the $4\times 2$-matrices $K_{i}$
and their $2\times 4$ hermitian conjugates $K_{i}^{+}$ defined by $\cite%
{Belinfante}$%
\begin{eqnarray}
K_{1}^{+} &=&\hbar \left( 
\begin{array}{cccc}
-\sqrt{\frac{3}{2}} & 0 & \sqrt{\frac{1}{2}} & 0 \\ 
0 & -\sqrt{\frac{1}{2}} & 0 & \sqrt{\frac{3}{2}}%
\end{array}%
\right) ,K_{2}^{+}=\hbar \left( 
\begin{array}{cccc}
-i\sqrt{\frac{3}{2}} & 0 & -i\sqrt{\frac{1}{2}} & 0 \\ 
0 & -i\sqrt{\frac{1}{2}} & 0 & -i\sqrt{\frac{3}{2}}%
\end{array}%
\right) ,  \label{51} \\
K_{3}^{+} &=&\hbar \left( 
\begin{array}{cccc}
0 & \sqrt{2} & 0 & 0 \\ 
0 & 0 & \sqrt{2} & 0%
\end{array}%
\right) .  \notag
\end{eqnarray}%
The $K_{i}$ and $s_{i}$ matrices and the components of the the momentum $%
\boldsymbol{p}$ obey the relation%
\begin{equation}
\left( 4/9\right) \hbar ^{-2}\left( K_{j}K_{i}^{+}+s_{j}s_{i}\right)
p_{j}p_{i}=p^{2}1_{4\times 4}.  \label{52}
\end{equation}%
Since the previous equation is similar to $\left( \ref{28}\right) $, one can
take the form given by $\left( \ref{32}\right) $ where $N_{i}$ and $\hbar $
are replaced by $K_{i}$ and $3\hbar /2$. As in the spin one case, we find
that the $\left( 6s+1\right) \times \left( 6s+1\right) $ following matrix%
\begin{equation}
B_{j}B_{i}p_{j}p_{i}=\left( 4/9\right) \hbar ^{-2}\left( 
\begin{array}{ccc}
K_{j}K_{i}^{+}+s_{j}s_{i} & 0 & 0 \\ 
0 & K_{j}^{+}K_{i} & K_{j}^{+}s_{i} \\ 
0 & s_{j}K_{i} & s_{j}s_{i}%
\end{array}%
\right) p_{j}p_{i},  \label{54}
\end{equation}%
is not equal to $p^{2}$ and therefore $\left( \ref{8}\right) $ is not
satisfied. Again, the requirement that the wave function must be taken into
account leads to $\left( \ref{35}\right) $, where now $\psi $ is a
ten-component object

\begin{equation}
\psi =\left( 
\begin{array}{c}
\varphi  \\ 
\Omega  \\ 
\chi 
\end{array}%
\right) ,  \label{55}
\end{equation}%
here $\varphi $ and $\chi $ are four-component and $\Omega $ is a
two-component functions. Equation $\left( \ref{35}\right) $ yields for the
spin $3/2$ particle%
\begin{equation}
\left( 4/9\right) \hbar ^{-2}\left[ \left( K_{j}p_{j}\right) \left(
K_{i}^{+}p_{i}\right) +\left( s_{j}p_{j}\right) \left( s_{i}p_{i}\right) %
\right] \varphi =p^{2}\varphi ,  \label{56}
\end{equation}%
\begin{equation}
\left( 4/9\right) \hbar ^{-2}\left( K_{j}^{+}p_{j}\right) \left[ \left(
K_{i}p_{i}\right) \Omega +\left( s_{i}p_{i}\right) \chi \right] =p^{2}\Omega
,  \label{57}
\end{equation}%
\begin{equation}
\left( 4/9\right) \hbar ^{-2}\left( s_{j}p_{j}\right) \left[ \left(
K_{i}p_{i}\right) \Omega +\left( s_{i}p_{i}\right) \chi \right] =p^{2}\chi .
\label{58}
\end{equation}%
The first equation is obvious and the last two equations leads to conditions
on the components of the wave function%
\begin{equation}
\Omega =-\left( 2\alpha /3\right) \left( K_{i}^{+}p_{i}\right) \varphi 
\label{59}
\end{equation}%
\begin{equation}
\chi =-\left( 2\alpha /3\right) \left( s_{i}p_{i}\right) \varphi ,
\label{60}
\end{equation}%
where $\alpha $ is a constant to be determined. The last step is to
construct the $A$\ and $C$\ matrices consistent with the two previous\
equations. We proceed with a method different to the one used in the
previous cases and without using $\left( \ref{6}\right) $ to $\left( \ref{8}%
\right) $ deduced from the linearization procedure. Taking into account the
expression of $B_{i}$ and $\left( \ref{55}\right) $, the resolution of
equation%
\begin{equation}
B_{i}p_{i}\psi =-(AE+C)\psi ,  \label{61}
\end{equation}%
subject to the constraints $\left( \ref{59}\right) $, $\left( \ref{60}%
\right) $ and $\left( \ref{4}\right) $, we find that $A$\ and $C$ are given
by expressions similar to that of spin one case with the appropriates
dimensions and the $\alpha $\ constant is given by $\alpha =(2m\hbar )^{-1}$%
. Finally, the $\left( 6s+1\right) $ components first-order wave equation
for this particle is given by%
\begin{equation}
\left[ \left( 
\begin{array}{ccc}
1 & 0 & 0 \\ 
0 & 0 & 0 \\ 
0 & 0 & 0%
\end{array}%
\right) E+\left( 3\hbar /2\right) ^{-1}\left( 
\begin{array}{ccc}
0 & \boldsymbol{K} & \boldsymbol{s} \\ 
\boldsymbol{K}^{+} & 0 & 0 \\ 
\boldsymbol{s} & 0 & 0%
\end{array}%
\right) \boldsymbol{p}+2m\left( 
\begin{array}{ccc}
0 & 0 & 0 \\ 
0 & 1 & 0 \\ 
0 & 0 & 1%
\end{array}%
\right) \right] \psi =0.  \label{62}
\end{equation}%
This result coincides with the Hurley's equation $\cite{Hurley}$ for spin
3/2 particle.

For the harmonic oscillator problem, again we perform the substitution $%
\boldsymbol{p\rightarrow p}-im\omega \eta \boldsymbol{r}$ with $\eta
=2A^{2}-1$. The decomposition of the wave aquation into its components and
the elimination of $\Omega $\ and $\chi $\ in favour of $\varphi $\ gives%
\begin{equation}
2mE\varphi =\left( 4/9\right) \hbar ^{-2}\left[ \left( \boldsymbol{K}.%
\boldsymbol{p}_{+}\right) \left( \boldsymbol{K}^{+}.\boldsymbol{p}%
_{-}\right) +\left( \boldsymbol{s}.\boldsymbol{p}_{+}\right) \left( 
\boldsymbol{s}.\boldsymbol{p}_{-}\right) \right] \varphi   \label{63}
\end{equation}%
with the help of the property $\cite{Belinfante}$%
\begin{equation}
\boldsymbol{K}_{i}\boldsymbol{K}_{j}^{+}+s_{i}s_{j}=i\left( 3\hbar /2\right)
\varepsilon _{ijk}s_{k}+\left( 9\hbar /4\right) \delta _{ij}
\end{equation}%
\ obtained using the definitions $\left( \ref{50}\right) $ and $\left( \ref%
{51}\right) $, the evaluation of the right-hand side of $\left( \ref{63}%
\right) $\ yields%
\begin{equation}
E\varphi =\left[ \frac{p^{2}}{2m}+\frac{1}{2}m\omega ^{2}r^{2}-\frac{3}{2}%
\hbar \omega -\frac{2\omega }{3\hbar }\left( \boldsymbol{L}.\boldsymbol{s}%
\right) \right] \varphi ,
\end{equation}%
which corresponds to the isotropic harmonic oscillator plus the spin-orbit
coupling whose strength is one third to the one obtained for spin 1/2 case.

\section{Conclusion}

A new method is presented based on the linearization of the wave equation
which allow us to deduce the first order equations for non-relativistic
particles of spin 1/2, 1 and 3/2. Requiring that the wave function $\psi $
must be taken into account in the set of relations $\left( \ref{6}\right) $
to $\left( \ref{8}\right) $, we were able to construct the operators $A$ , $%
B_{i}$, $C$ for particles of spin 1 and spin 3/2. This requirement is no
longer necessary for a particle of spin 1/2 and the resolution of the
previous set of equations leads then to the particular solution established
by L\'{e}vy-Leblond.

In addition, we considered the problem of the harmonic oscillator in the
framework of linear wave equations and we were able to find the standard
wave equations plus the terms giving the spin-orbit coupling in each case.

The method presented in this paper can be generalized to the case of a
non-relativistic (and relativistic) particle of arbitrary spin as well as
that of a spinless particle (works in progress).

\end{document}